# A Raman-type optical frequency comb adiabatically generated in an enhancement cavity

*M. Katsuragawa[1,2], R. Tanaka[1], H. Yokota[1], and T. Matsuzawa[1]

[1]*Department of Applied Physics and Chemistry, University of Electro-Communications,
1-5-1 Chofugaoka, Chofu, Tokyo 182-8585, Japan*

[2] *PRESTO, Japan Science and Technology Agency,
4-1-8 Honcho, Kawaguchi, Saitama, Japan*

**Abstract**

We demonstrate efficient generation of a Raman-type optical frequency comb by employing adiabatic Raman excitation in an enhancement cavity. A broad frequency comb spanning over 130 THz is realized with an excitation power reduction exceeding three orders of magnitude compared with a single-pass configuration.

A method that drives a nonlinear optical process in a cavity has been employed as a key technique for overcoming the weak field power that prevents the realization of an efficient nonlinear optical process[1]. This method was first demonstrated 40 years ago. It realized the second harmonic generation of continuous wave (cw) He-Ne laser radiation[2]. In the 1980's, the technique of frequency locking to a cavity was introduced, and thus single-frequency second harmonic generation was achieved[3]. A recent significant study would be the generation of a high-harmonic comb in the vacuum ultraviolet region that was realized by confining a femtosecond mode-locked laser in a high finesse cavity[4,5]. Here, we demonstrate adiabatic generation of frequency comb by placing a Raman medium in an enhancement cavity and confining nanosecond pulses in it.

Figure 1a is a conceptual schematic of this study. A pair of single frequency nanosecond pulses whose difference frequency is near resonant to a Raman transition is produced from a single laser resonator. We confine these nanosecond pulses stably in an enhancement cavity filled with a Raman medium, and then adiabatically drive parametric stimulated Raman scattering in the enhancement cavity.

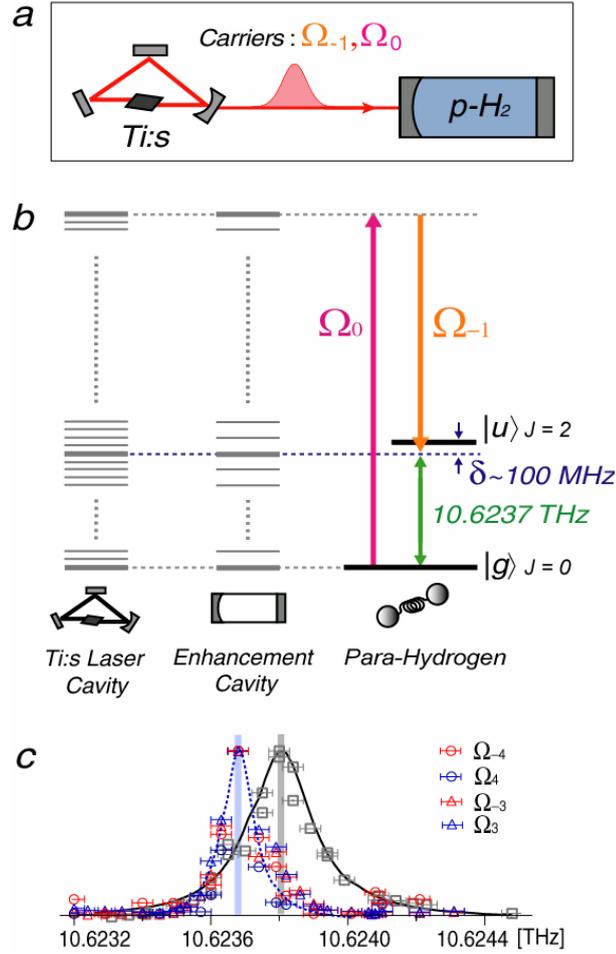

**Fig. 1: Scheme for adiabatic Raman sideband generation in an enhancement cavity.** *a*, Conceptual illustration. *b*, Detailed configuration for Raman sideband generation in an enhancement cavity. The following must satisfy triple resonance conditions, the Ti:sapphire (Ti:S) laser cavity, the enhancement cavity, and optimum two-photon detuning for the adiabatic excitation of Raman coherence. *c*, Typical example depicting optimal two-photon detuning without the enhancement cavity. The black curve is a resonance profile of the rotational Raman transition (J = 2 <- 0) in parahydrogen. The circles and triangles indicate the generated intensities of ±3 and ±4 sidebands (normalized), shown as a function of two-photon detuning.

The right panel in Fig. 1b shows the scheme of the adiabatic parametric stimulated Raman scattering. The states of |g> and |u> form a two-level system in which a Raman transition is allowed. We apply a pair of single-frequency pulses, $\Omega_0$ and $\Omega_{-1}$, which are near resonant to these two levels, and drive this Raman system from the initial state, |g>. When the two-photon detuning $\delta$, is set so that the Raman transition and the difference frequency of the excitation laser pulses, including their spectral broadenings, $\delta\nu_m$ and

$\delta\nu_L$, respectively, do not overlap, i.e., $\delta > 1/2\ (\delta\nu_m + \delta\nu_L)$, the excitation process becomes adiabatic[6,7]. When we control $\delta$ at the minimum values that satisfy this adiabatic condition, i.e., $|\delta_{opt}| \sim 1/2\ (\delta\nu_m + \delta\nu_L)$, a realistic excitation power can satisfy $\Omega_{gu}/\delta_{opt} \gg 1$ to produce the theoretical-limit coherence, $|\rho_{gu}| = 0.5$, where $\Omega_{gu}$ and $\rho_{gu}$ are the two-photon Rabi frequency and the off-diagonal term of the density matrix expressing the two-level system of $|g\rangle$ and $|u\rangle$[6,7].

The resulting molecular oscillation with very high coherence, in turn, deeply modulates the two excitation pulses, and generates a high order series of parametric stimulated Raman scatterings (we call these 'Raman sidebands')[6-9]. A noteworthy feature of this Raman sideband generation is that efficient generation can occur within a unit phase-slip length, since the produced coherence is extremely high. This results in coaxial generation of the sidebands over a wide spectral range, without any restriction being imposed by the phase matching condition[6-9].

Figure 1c shows a case employing a pure rotational transition of *J = 2 <- 0* in parahydrogen (density: $4.8 \times 10^{19}$ cm$^{-3}$, Raman transition frequency: 10.62380 THz, spectral width: 110 MHz at HWHM)[10]. The highest coherence is produced at a two-photon detuning of + 120 MHz from the exact Raman resonance, corresponding approximately to $1/2\ (\delta\nu_m + \delta\nu_L)$, and thereby the coaxial high-order sidebands ($\Omega_{\pm 3}$: ±3 orders, $\Omega_{\pm 4}$: ±4 orders) are efficiently generated.

In this study, we carry out this adiabatic Raman sideband generation inside an enhancement cavity. The concept is already outlined in Fig. 1a. A more quantitative configuration is shown in Fig. 1b. Both of the excitation nanosecond pulses, $\Omega_0$ and $\Omega_{-1}$, which provide the optimal two-photon detuning for the adiabatic Raman excitation, must simultaneously match each longitudinal mode of the excitation laser (free spectral range (FSR): 1.24 GHz) and the enhancement cavity (FSR: 1.98 GHz); namely the triple resonant condition, so as to be stably confined in the enhancement cavity.

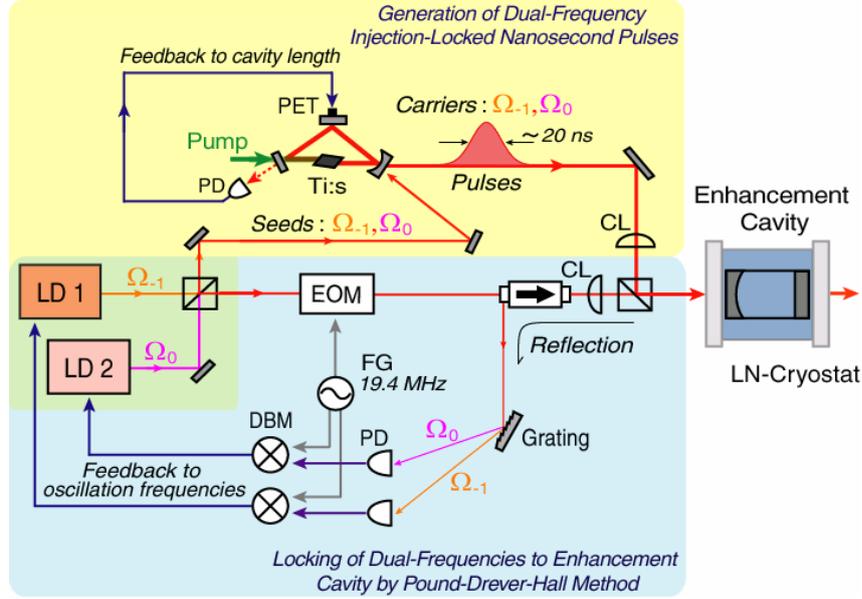

**Fig. 2: Schematic illustrating the experimental setup for Raman sideband generation in an enhancement cavity.** LD, laser diode, EOM, electrooptic modulator, FG, function generator, DBM, double balanced mixer, PD, photo diode, PET, piezoelectric transducer, CL, coupling lens, LN-cryostat, liquid nitrogen cryostat.

Figure 2 is a conceptual illustration of the experimental setup. The complete system consists of three parts: a system for generating cw laser radiation at two frequencies stabilized to the enhancement cavity (highlighted in blue), a system for generating nanosecond pulses at two frequencies as carriers (highlighted in yellow), and an enhancement cavity filled with gaseous parahydrogen.

The simultaneous confinement of the nanosecond pulses at two frequencies in the enhancement cavity was realized as follows. LD1($\Omega_{-1}$: 808 ± 3 nm) and LD2($\Omega_0$: 785 ± 3 nm) are external-cavity-controlled laser diodes. Each oscillation frequency was simultaneously stabilized to the enhancement cavity by employing two independent feedback loops based on the Pound-Drever-Hall (PDH) method[11]. These cw laser radiations were also extracted from the other outlet of the cubic beam splitter and introduced into the injection-locked nanosecond pulsed Ti:S laser[12, 13] as seeds (typically 1 mW). By referencing these two seed frequencies, we finely adjusted the Ti:S cavity length and stabilized it in a doubly resonant condition with a mismatch of less than a few MHz. The mismatch between the resonant conditions of the two seed frequencies was compensated by tuning the cavity length of the Ti:S laser itself. The compensation amounted to 1/36 FSR for every 1 FSR cavity-length change. In the doubly resonant condition, the pulsed outputs were simultaneously injection-locked to the two seeds. Thus the nanosecond pulses were produced with carrier frequencies, $\Omega_{-1}$ and $\Omega_0$, both stabilized to the enhancement cavity. Finally, the nanosecond pulses were transformed so as to match the transverse mode of the enhancement cavity, and coupled into it.

We employed a Fabry-Perot configuration for the enhancement cavity. It consisted of a pair of concave (r = 200 mm) and flat mirrors, each with a reflectivity of 98.75% for a 730 - 900 nm spectral range. The cavity length, FSR, and finesse were 75.00 mm, 2.00 GHz, and ~ 250, respectively. The waist diameter of the fundamental transverse mode of this cavity, $TEM_{00}$, was 0.31 mm. This corresponds to a Rayleigh length of 94 mm, provided a Gaussian beam in free space. The enhancement cavity was installed in a sample cell and placed in a liquid nitrogen cryostat to realize Raman sideband generation at the optimal temperature (50 - 100 K). The sample cell was filled with pure parahydrogen (> 99.9%) [14], and its density was controlled in the $(2.9 \sim 0.48) \times 10^{20}\,cm^{-3}$ range.

To match the optimum condition needed for the adiabatic Raman process with the requirements for the enhancement cavity sideband generation (Fig. 1b), we systematically investigated the spectroscopic properties of the target rotational Raman transition in advance, namely the density dependence of the resonant frequency and the decoherence time. Based on this fundamental data, we designed and manufactured the enhancement cavity length to the target value with a precision of better than 10 μm. The design included the effects of the refractive index of parahydrogen and the thermal shrinking of the cavity length from room temperature to liquid nitrogen temperature. Finally, around the target densities, we adjusted the parahydrogen density slightly (a few %) to achieve fine control to the optimal adiabatic conditions.

Raman sideband generation was carried out in the enhancement cavity around the triple resonant condition shown in Fig. 1b. The seed laser frequencies were monitored with a wavemeter (Burleigh, WA1500, resolution: $10^{-7}$). The spectrum and temporal profiles of the generated Raman sidebands were measured with an optical multichannel analyzer (OMA) and biplanar phototubes (Hamamatsu Photonics K.K., R1328U, build-up time: 60 ps), respectively.

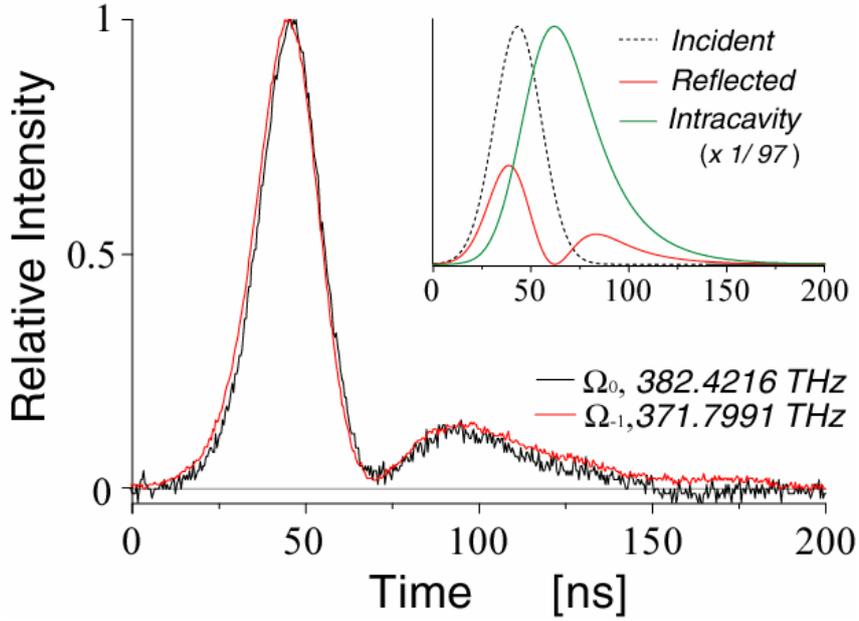

**Fig. 3: Confinement of two-frequency excitation nanosecond pulses in an enhancement cavity.** The reflected waveforms are shown for the two excitation pulses, $\Omega_0$ (black), $\Omega_{-1}$ (red). The inset depicts the reflected and intracavity waveforms calculated for the incident pulse (duration: 25 ns at FWHM).

Figure 3 shows the reflected waveforms of the excitation laser pulses introduced into the enhancement cavity (black: $\Omega_0$, 382.4216 THz, 783.9319 nm; red: $\Omega_{-1}$, 371.7991 THz, 806.3292 nm). Both incident pulses had durations of 25 ns at full width at half maximum (FWHM). The pulse energies were set to be so weak that they did not generate any Raman sidebands. It is clearly seen that the reflected waveforms are identical for the two-frequency pulses. They have two peaks and a dip that approaches zero between the two peaks. The shapes of their waveforms were stable shot-by-shot. The inset shows the waveforms calculated using the conditions employed for the experiment. The black dots, and red and green solid curves correspond to incident, reflected, and intracavity waveforms, respectively[15]. The observed reflected waveform agrees well with the expected one. Based on the system shown in Fig. 2, the two-frequency excitation pulses were simultaneously and stably confined in the enhancement cavity.

As depicted in the inset, here the intracavity radiation forms a waveform appropriate for the adiabatic excitation that smoothly builds up and decreases with the cavity lifetime (pulse duration, 43 ns at FWHM). Furthermore, due to this confinement, the peak power of the intracavity waveform is 97 times greater than that of the incident excitation pulse.

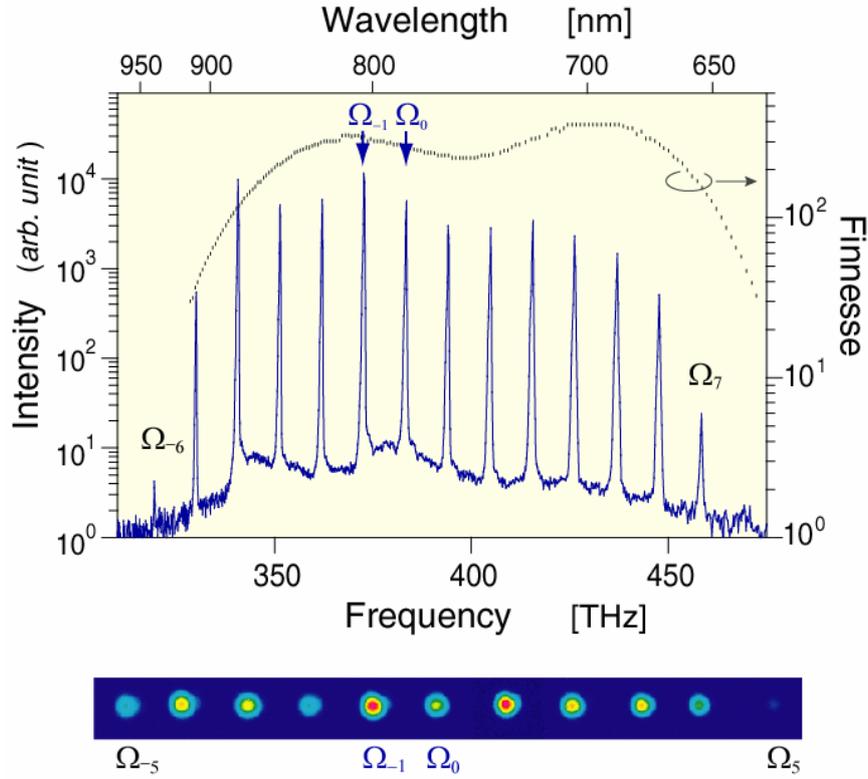

**Fig. 4: Raman sidebands adiabatically generated in enhancement cavity.** The density of parahydrogen and the excitation power introduced into the enhancement cavity were 4.8 x $10^{19}$ cm$^{-3}$ and 0.27 kW, respectively. For the sideband spectrum above, the two-photon detuning was set at + 80 MHz. The dotted curve represents the frequency dependence of the finesse of the enhancement cavity. Typical beam profiles of the Raman sidebands are shown below, where the two-photon detuning was set at + 180 MHz.

In the density control range, (2.9 ~ 0.48) x $10^{20}$ cm$^{-3}$, we were able to select six different conditions that satisfied the triple resonances shown in Fig. 1b. At the respective conditions, we raised the excitation power, and performed Raman sideband generation with the enhancement cavity. The sidebands were observed both forward and backward and they were similar. The dependence on the two-photon detuning was nearly the same as that found in a single-pass configuration. On the other hand, the dependence on the parahydrogen density was rather different. As expected, more efficient sidebands were observed for lower densities at which the optimum two-photon detuning could be small, while in a single-pass configuration, it was nearly independent of density.

Figure 4 is a typical Raman sideband spectrum observed with the enhancement-cavity configuration. The medium density was set at the lowest value that satisfied the triple resonance condition (4.8 x $10^{19}$ cm$^{-3}$). Here, the two-photon detuning was set at + 80 MHz ($\Omega_0$ - $\Omega_{-1}$: 10.62372 THz). This value was almost in accordance with the optimum two-photon detuning of 120 MHz found in a single-pass configuration (Fig. 1c).

Fourteen sidebands are clearly seen that spread over a broad spectral range of 656 – 941 nm ($\Omega_{+7}$: 656.32 nm, 456.78 THz – $\Omega_{-6}$: 940.73 nm, 318.69 THz). The sidebands have an equidistant frequency spacing that is near resonant to the molecular Raman transition. This is like a Raman-type optical frequency comb. Although a very broad spectrum was produced, the peak power of each excitation laser was only 0.27 kW (energy: 6.8 μJ, duration: 25 ns). This is three orders of magnitude less than that of a typical single-pass configuration.[16, 17]

The dotted curve in Fig. 4 represents the finesse of the employed enhancement cavity. The nearly flat intensity distribution of the generated sidebands well reflects behavior similar to the spectral dependence of the finesse. This is a manifestation of the typical aspect that the confinement in the enhancement cavity dominates the sideband generation process.

The lower panel in Fig. 4 shows typical beam profiles of the generated Raman sidebands (two-photon detuning: 180 MHz). These were measured with a charge coupled device-based beam profiler. We see that all the sideband components have round, high quality beam profiles reflecting the fundamental transverse mode (TEM$_{00}$) of the enhancement cavity.

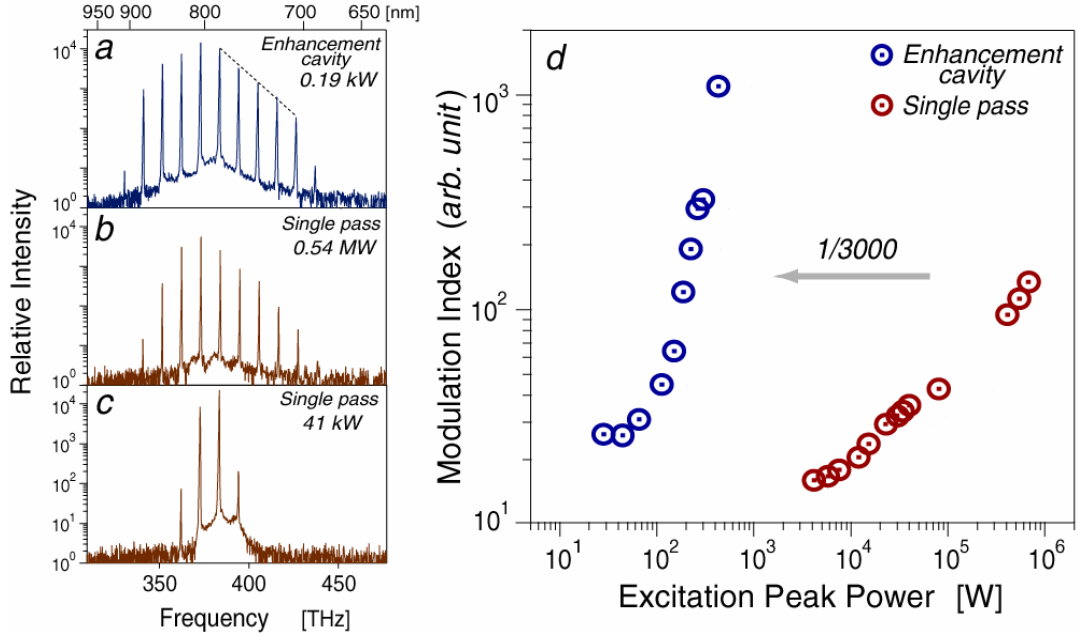

**Fig. 5: Raman sideband generation, Enhancement cavity vs. Single pass.** The left panel shows the sideband spectra for a, enhancement cavity (excitation power, 0.19 kW), b, single pass (0.54 MW), and c, single pass (41 kW). The right panel shows modulation indexes, estimated from the generated sideband spectra as a function of the excitation power, for enhancement cavity (blue circles) and single-pass (brown circles) configurations.

Finally, in Fig. 5, we show the sideband generation dependence on excitation power, which focuses on a comparison with a single-pass configuration. Panel 5a is a sideband spectrum with the enhancement cavity configuration, obtained at an excitation power of 0.19 kW (4.7 µJ, 25 ns). We also find a similar spectrum in Fig. 5b, but this was obtained with a single-pass configuration. The excitation power was 0.54 MW (3.5 mJ, 6.5 ns), which is 2800 times more intense than that in 5a. If we reduce the excitation power by one order of magnitude to 41 kW (260 µJ, 6.5 ns) as shown in Fig. 5c, only ± 1 orders of sidebands were seen. The higher orders were not generated at all. We find that the sidebands in 5a with the enhancement cavity were realized at an excitation power 200 times further lower than those in 5b.

The generation process of the Raman sidebands is analogous to optical phase modulation using an electrooptic modulator. That is, the inclination of the intensity distribution of the Raman sidebands, expressed in a log scale, can be in inverse proportion to the product of the phase modulation index, $\beta$, and the cavity finesses, $F$, $\beta F$ [18]. Figure 5d plots the estimated inclinations as a function of the excitation power. This is examined by focusing on the high frequency (anti-Stokes) side. An example is shown by the dotted line in Fig. 5a. It is clearly seen that due to the enhancement cavity generation, the required excitation power is certainly reduced by three orders (~ 1 / 3000).

In the enhancement cavity configuration employed in this study, the excitation laser power ($\propto \beta$) can be enhanced 96.7 times. We can consider the effective finesse ($F$) for the nanosecond excitation pulses to be similar. Using these values, when we briefly estimate the Raman sideband generation in the enhancement cavity, we can expect a reduction factor of 9,350 ($\beta F$ ~ 96.7 x 96.7) for the required excitation power. The result in Fig. 5d is reasonably consistent with this brief estimation.

In summary, we have demonstrated extremely efficient Raman sideband generation (Raman type optical frequency comb) based on adiabatic Raman excitation in an enhancement cavity. We have shown that optimization for both of the adiabatic-excitation condition in the enhancement cavity and the medium temperature and density, realized broad Raman sidebands with a nearly flat intensity distribution over 130 THz (656 – 941 nm). We have also shown that this enhancement cavity technique provides an excitation power reduction exceeding three orders of magnitude compared with a single-pass configuration.

The present study can be extended to quasi-cw and/or cw regimes[19, 20]. This would be very attractive as a novel method for realizing an optical frequency comb originating from single-frequency lasers[21-23], and also for generating arbitrary optical waves[24] with an ultrahigh frequency repetition rate. We thank K. Hakuta for valuable comments. We also thank T. Suzuki and A. Mio for helpful discussions and support with the experiments.

**References**

1. Y. R. Shen, *The Principle of Nonlinear Optics* (Wiley-Interscience, New Jersey, 2003).; R.W. Boyd, *Nonlinear Optics* (Academic, San Diego, 2003).
2. A. Ashkin, G. D. Boyd, and J. M. Dziedzic, *IEEE J. Quantum Electron.*, **QE-2**, 109 (1966).
3. M. Brieger, *et al.*, *Opt. Commun.* **38**, 423 (1981).
4. C. Gohle, *et al.*, *Nature* **436**, 234 (2005).
5. R. J. Jones, *et al.*, *Phys. Rev. Lett.* **94**, 193201 (2005).
6. S. E. Harris and A. V. Sokolov, *Phys. Rev. A* **55**, R4019 (1997).
7. L. K. Fam, *et al. Phys. Rev. A* **60**, 1562 (1999).
8. A. V. Sokolov, *et al. Phys. Rev. Lett.* **85**, 562 (2000).
9. J. Q. Liang, *et al.*, *Phys. Rev. Lett.* **85**, 2474 (2000).
10. This spectral width includes those of the lasers (~ 13 MHz) used in this measurement, implying to be $1/2\ (\delta\nu_m + \delta\nu_L)$.
11. R. W. P. Drever, et al., *Appl. Phys. B: Lasers and Optics*, **31**, 97 (1983).
12. M. Katsuragawa, and T. Onose, *Opt. Lett.* **30**, 2421 (2005).
13. T. Onose, and M. Katsuragawa, *Opt. Exp.* **15**, 1600 (2007).
14. M. Suzuki, *et al.*, *J. Low Temp. Phys.* **111**, 463 (1998).
15. R. Tanaka, *et al.*, *Opt. Exp.* **16**, 18667 (2008).
16. M. Katsuragawa, *et al.*, *Opt. Exp.* **13**, 5628 (2005).; M. Katsuragawa, *et al.*, CLEO/QELS 2006, QELS Technical Digest, QFE1, Long Beach, California, USA, May. 21-26 (2006).
17. T. Suzuki, M. Hirai, and M. Katsuragawa, *Phys. Rev. Lett.* **101**, 243602 (2008).
18. M. Kourogi, K. Nakagawa, and M. Ohtsu, *IEEE J. Quantum Electron.* **29**, 2693 (1993).
19. J. K. Brasseur, K. S. Repasky, and J. L. Carlsten, *Opt. Lett.* **23**, 367 (1998).
20. D. D. Yavuz, *Phys. Rev.* A **76**, 011805(R) (2007).
21. S. Uetake, *et al.*, *Phys. Rev. A* **61**, R 011803 (2000).
22. F. Benabid, *et al. Nature* **434**, 499 (2005).
23. P. Del'Haye, *et al. Nature* **450**, 1214 (2007).
24. Z. Jiang, *et al.*, *Nature Photon.* **1**, 463 (2007).